\documentclass[showpacs,preprintnumbers,amsmath,amssymb,prl]{revtex4}

\usepackage{graphicx}
\usepackage{bm}

\begin{document}

\title{Dynamical Crossover in Supercritical Core-Softened Fluids}

\author{Eu. A. Gaiduk}
\affiliation{Institute for High Pressure Physics, Russian Academy
of Sciences, Troitsk 142190, Moscow, Russia \\ Moscow Institute of
Physics and Technology, Dolgoprudny, Moscow Region 141700, Russia}
\author{Yu. D. Fomin}
\affiliation{Institute for High Pressure Physics, Russian Academy
of Sciences, Troitsk 142190, Moscow, Russia \\ Moscow Institute of
Physics and Technology, Dolgoprudny, Moscow Region 141700, Russia}
\author{V. N. Ryzhov}
\affiliation{Institute for High Pressure Physics, Russian Academy
of Sciences, Troitsk 142190, Moscow, Russia \\ Moscow Institute of
Physics and Technology, Dolgoprudny, Moscow Region 141700, Russia}
\author{E. N. Tsiok}
\affiliation{Institute for High Pressure Physics, Russian Academy
of Sciences, Troitsk 142190, Moscow, Russia}
\author{V. V. Brazhkin}
\affiliation{Institute for High Pressure Physics, Russian Academy
of Sciences, Troitsk 142190, Moscow, Russia}
\date{\today}

\begin{abstract}
It is well known that some liquids can demonstrate anomalous
behavior. Interestingly, this behavior can be qualitatively
reproduced with simple core-softened isotropic pair-potential
systems. Although anomalous properties of liquids usually take
place at low and moderate temperatures it was recently recognized
that many important phenomena can appear in supercritical fluids.
However, no studies of supercritical behavior of core-softened
fluids is reported. This paper reports a study of dynamical
crossover in supercritical core-softened systems. The crossover
line is calculated from three different criteria and good
agreement between them is observed. It is found that the behavior
of the dynamical crossover line of core-softened systems is quite
complex due to its quasi-binary nature.
\end{abstract}

 \maketitle


The most fundamental approach to the behavior of matter relies on
the interactions between the particles of the substance. One can
use an approach based on quantum mechanical treatment of
interactions which is the most accurate one. However, such
ab-initio methods require a lot of computational resources and
cannot be applied to the systems larger then several hundreds of
atoms. A great advantage was made by application of so called
effective potentials. These potentials are constructed in such a
way that they allow to obtain some principal properties of
interest with much smaller efforts. In the simplest case the
interaction energy is approximated by pair interactions only.
Among the most studied in the literature is Lennard-Jones (LJ)
system which has the potential $U(r)=\varepsilon \left( \left(
\frac{\sigma}{r} \right)^{12} - \left( \frac{\sigma}{r}
\right)^{6} \right)$. This system demonstrates a generic view of
phase diagram of a substance containing gas, liquid and crystal
phases and well describes the behavior of noble gases and some
molecular substances.

In our recent works it was shown that supercritical region of the
phase diagram can be divided into two parts: rigid liquid and
dense gas \cite{ufn,frpre,frprl}. These regions differ by the
microscopic dynamics of particles and are separated by a crossover
line called Frenkel line. Later on the phenomenon of dynamical
crossover was studied for a number of other fluids
\cite{frenkel-iron,wid-co2,kostyachina,fr-water,frenkel-h2}.


Another topic attracting wide attention of researchers is related
to anomalous behavior of liquids (see, for example,
\cite{wateranomalies} for the list of anomalies of water). It was
found that models with isotropic pair core-softened potentials can
demonstrate anomalous behavior
\cite{jcp-init,pre-init,attract-init,st,cs1,cs2,cs3,cs4}.
Diffusion, density and structural anomalies are widely discussed
in the literature \cite{buldyrev-rev}. Such systems can also
demonstrate numerous structural phase transitions in solid region.
This kind of behavior cannot be obtained in systems like LJ. These
observations allow to suppose that the behavior of the Frenkel
line can also be more complex in the systems with core-softened
potentials.

A particular form of core-softened system studied in our previous
works is characterized by the following interaction potential:

\begin{equation} \label{smooth-attr}
U(r)/ \varepsilon = \left(\frac{d}{r}\right)^{n}+ \lambda_0 -
\lambda_1 th(k_1[r- \sigma_1]) + \lambda_2 th(k_2 [r- \sigma_2]),
\end{equation}
where $d$ and $\varepsilon$ set the length and energy scales and
$\lambda_i$ and $\sigma_i$ are varied. A large set of parameters
was considered. It was found that the phase diagram and anomalous
behavior of the system strongly depend on the potential
parameters. An important feature of this system is that it
demonstrates quasi-binary behavior, i.e. some features commonly
observed in binary mixtures \cite{jcp-init}. For example, the
system can be easily vitrified \cite{jcp-init,glass-cs}. The phase
diagram of the system consits of low density close packed FCC
phase, high density FCC phase and a set of intermediate structures
which can be considered as close packing of the particles at large
length scale $\sigma_1$, close packed structure at small length
scale $d$ and a set of different structures in the region where
the competition between these length scales takes place.
Obviously, the interplay between two length scales affects the
liquid properties too. Therefore it is of interest to monitor the
phenomena of dynamical crossover in fluid for such a system.

The system with potential (~\ref{smooth-attr}) reproduces many
unusual properties: complex phase diagram with maximum on the
melting line and many different solid phases, anomalous density,
diffusivity and structure etc. Basing on this observation we
believe that investigation of the dynamical crossover in this
system will be helpful for understanding of the behavior of
supercritical water and other anomalous fluids.



\begin{figure}
\includegraphics[width=7cm, height=7cm]{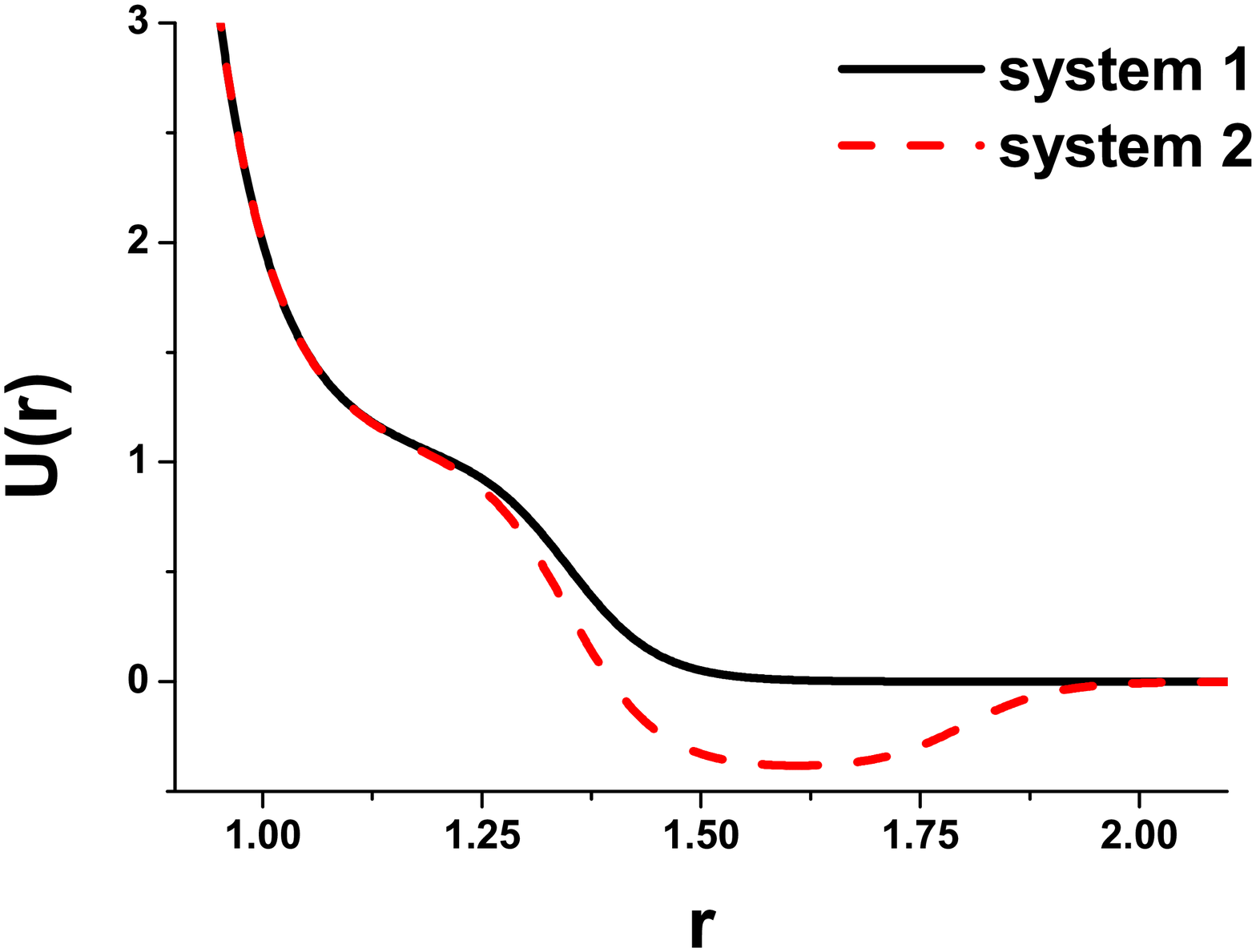}

\caption{\label{fig:fig1} (Color online) Interaction potentials
studied in the present work (eq. (\ref{smooth-attr})).}
\end{figure}

Two sets of parameter of the potential (\ref{smooth-attr}) were
studied: a purely repulsive system with the shoulder width
$\sigma_1=1.35$ and a system with repulsive shoulder and
attractive well. The parameters of the potentials are given in
Table I. Later on we refer the systems with different parameters
as "system 1" and "system 2". The potentials are shown in Fig.
~\ref{fig:fig1}.

We measure all properties of the system in reduced units with
respect to $d$ and $\varepsilon$: $\tilde{\rho}=N/V \cdot d^3$,
$\tilde{T}=T/(k_B \cdot \varepsilon)$ and so on. Since only these
reduced units are used in the paper, we omit the tilde marks.

In all cases systems of $1000$ particles in a cubic box with
periodic boundaries were simulated by molecular dynamics method.
The timestep was set to $0.0005$ reduced units of time. The system
was equilibrated by $3.5 \cdot 10^5$ followed by more $1.5 \cdot
10^5$ step for data collection. During the equilibration run the
temperature was held fixed by Nose-Hoover thermostat. The
propagation run was made in $NVE$ ensemble. All simulations were
performed with lammps simulation package \cite{lammps}.

\begin{table}

\begin{tabular}{|c|c|c|c|c|c|}
  \hline
   System number & $\sigma_1$ & $\sigma_2$ & $\lambda_0$ & $\lambda_1$ & $\lambda_2$ \\
  \hline
  1 & 1.35 & 0 & 0.5 & 0.5 & 0  \\
  \hline
  2 & 1.35 & 1.8 & 0.5 & 0.7 & 0.2  \\
  \hline
\end{tabular}

\caption{Parameters of the potential (\ref{smooth-attr}) used in
the study.}

\end{table}

Several criteria can be used to locate the line of dynamical
crossover \cite{ufn,frpre,frprl}. Among the most convenient ones
are the velocity autocorrelation function (vacf) criterion and the
isochoric heat capacity one \cite{frprl}.

Vacfs are defined as $Z(t)=\frac{1}{3N} \langle \sum \frac{{\bf
V}_{i}(t) {\bf V}_i(0)}{{\bf V}_i(0)^2} \rangle $ where ${\bf
V}_i(t)$ is an $i$-th particle velocity at time $t$. Vacfs of
rigid fluids demonstrate oscillatory behavior whilst vacfs of
dense gas decay monotonically. Therefore the Frenkel line
corresponds to the ($\rho,T$) points where the oscillations vanish
\cite{frprl}.

The isochoric heat capacity criterion states that at the Frenkel
line $c_v=2k_B$ where $c_v$ is heat capacity per particle. It is
based on the counting of contribution to the heat capacity of
liquid from longitudinal and transverse excitations. A detailed
theory of liquid heat capacity based on excitation spectra was
proposed in recent works \cite{phonon1,phonon2,scirep}. One can
show that in rigid regime the contribution to the heat capacity
from the potential energy of transverse modes is $1 \cdot k_B$ per
particle. At the Frenkel line transverse excitations disappear and
therefore the heat capacity per particle at Frenkel line should be
close to $2k_B$.

The crossover between different regimes of fluid also can be
observed by appearance of strong positive sound dispersion (PSD).
PSD means that the speed of excitations in liquids at some finite
wavelength $k$ exceeds the adiabatic speed of sound $c_s$. PSD was
experimentally observed in a number of low-temperature fluids
(see, for example, review \cite{single,sn,hoso3} and references
therein). As the temperature increases PSD disappears. Previously
we considered PSD in Lennard-Jones and soft spheres systems
\cite{frprl} and found that in both cases the temperature of PSD
loss is consistent with Frenkel temperature from vacf and
$c_v=2k_B$ criteria. The PSD disappearance is indeed related to
changes of excitation spectrum which take place at the Frenkel
line \cite{frprl}. Therefore it looks reasonable to relate the
disappearance of PSD with the crossing of the Frenkel line.


To check the presence or absence of PSD we compute the
longitudinal autocorrelation function of the velocity current
function:

\begin{equation}
C_L(k,t)=\frac{k^2}{N} <J_z({\bf k},t) \cdot J_z(-{\bf k},t)>,
\end{equation}
where

\begin{equation}
J({\bf k},t)=\sum_{j=1}^N {\bf v}_j e^{-i{\bf kr}_j(t)}.
\end{equation}

After calculation of $C_L({\bf k},t)$ we evaluated its Fourier
transform $\tilde{C}_L({\bf k},\omega)$. The frequency of the
excitations at wave vector ${\bf k}$ is given by the location of
the maximum of $\tilde{C}_L({\bf k},\omega)$.

Adiabatic speed of sound is calculated directly from simulations.



\bigskip

In the present work we calculate the Frenkel line from vacf,
$c_v=2k_B$ and PSD criteria for the system 1 and from vacf and
$c_v=2k_B$ criteria for the system 2.

\begin{figure}
\includegraphics[width=7cm, height=7cm]{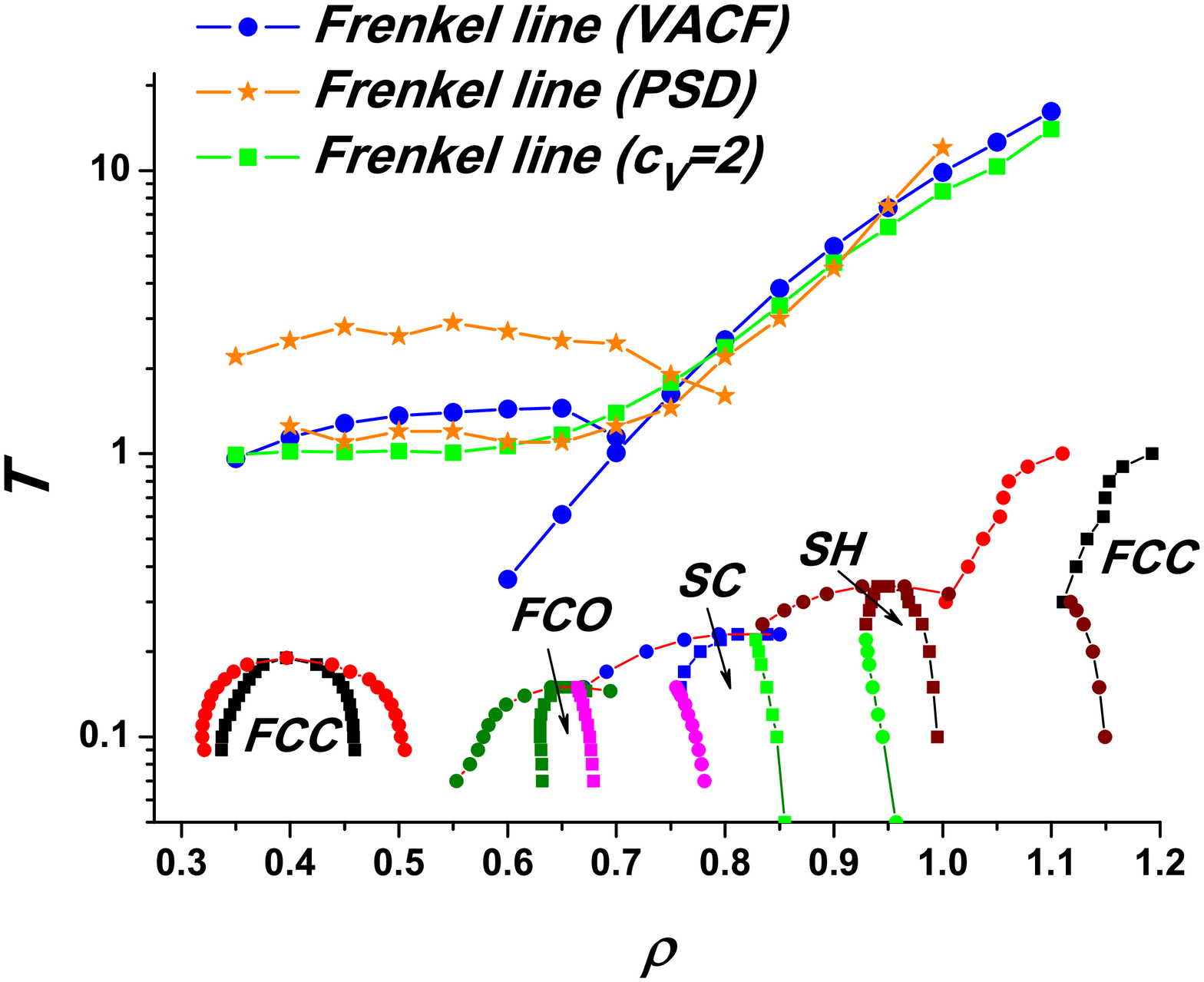}

\caption{\label{fig:fig4} (Color online) Frenkel line of the
system 1 computed from vacfs, $c_V=2k_B$ and PSD criteria placed
in the phase diagram.}
\end{figure}

Let us consider the Frenkel line of the system 1. We start with
calculation of the points where $c_v=2k_B$. The location of these
points in the phase diagram is shown in Fig.~\ref{fig:fig4}.

\begin{figure}
\includegraphics[width=7cm, height=7cm]{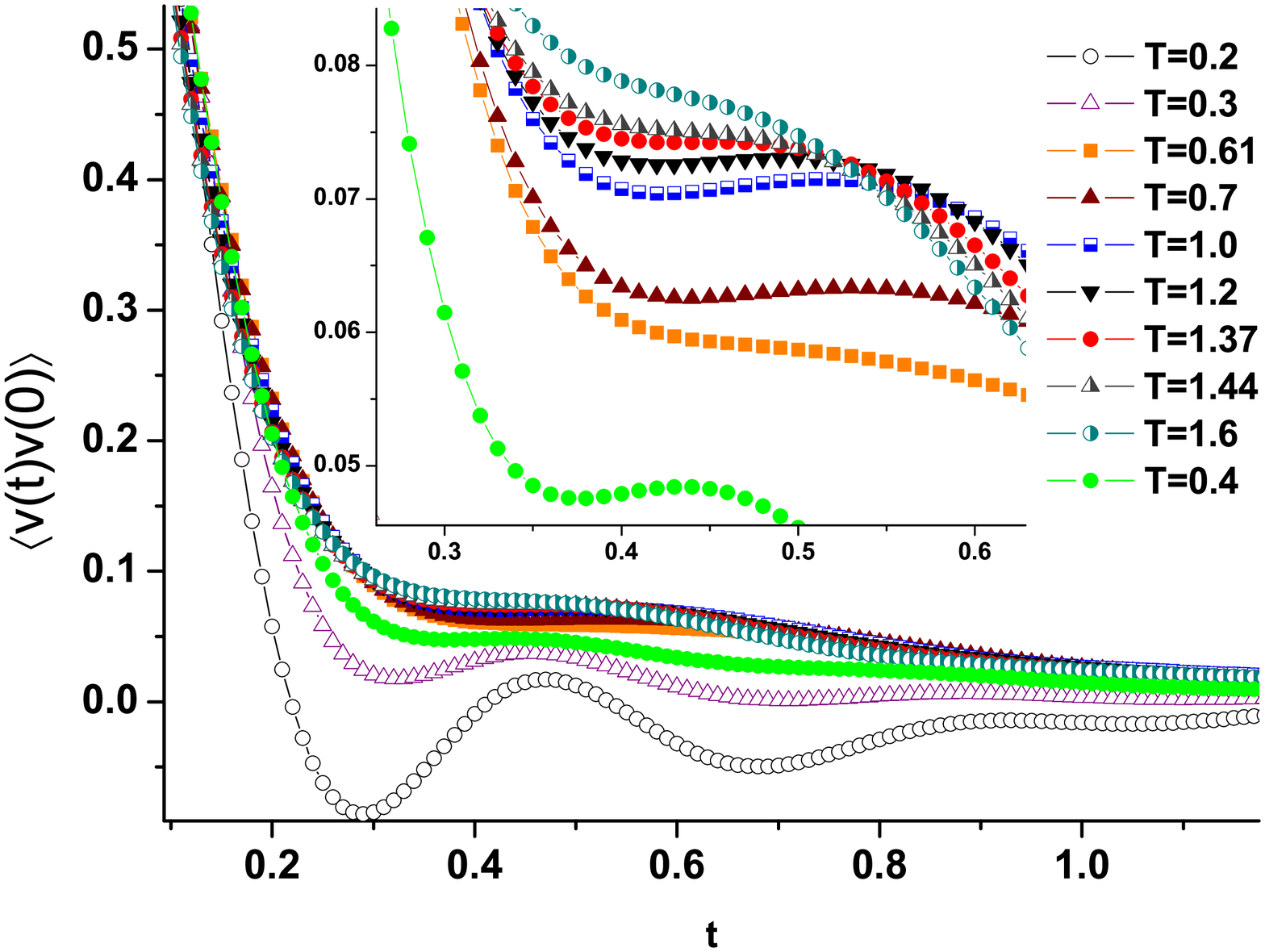}

\caption{\label{fig:fig2} (Color online) Vacfs of system 1 at
density $\rho =0.65$ and a setof temperatures. The inset enlarges
the region where the loss of oscillatory behavior takes place.}
\end{figure}

We proceed the study with calculation of the vacfs of the system
1. Fig.~\ref{fig:fig2} shows the evolution of vacfs of the system
at density $\rho=0.65$ as a function of temperature. One can see
strong oscillations at the lowest temperature. However, these
oscillations quickly decrease with temperature and disappear at
$T=0.61$. Interestingly, on further temperature increase the
oscillations of vacf appear again and disappear for the second
time at $T=1.45$. The points where vacfs loose theirs oscillations
are shown in Fig.~\ref{fig:fig4}.

This kind of reversible oscillation behavior of vacf is reported
for the first time. One can relate it to the quasibinary nature of
the system \cite{jcp-init}. The potential of the system 1 contains
two length scales: $d$ and $\sigma_1$. Such systems behave
qualitatively different at low and high temperature
\cite{jcp-init}. Since the potential is purely repulsive it is
energetically advantageous for the particles to stay far from each
other. Therefore at low temperatures and densities the behavior of
the system is defined by the parameter $\sigma_1$. If the density
and temperature increase the particles penetrate through the soft
core and the system behaves more likely as defined by the small
length scale $d$. Apparently, an intermediate region appears in
between in which the influence of both length scales is comparable
and therefore competition between local structures takes place.
This quasi-binary nature is obviously responsible for the complex
behavior of the vacfs of the system.

\begin{figure}
\includegraphics[width=7cm, height=7cm]{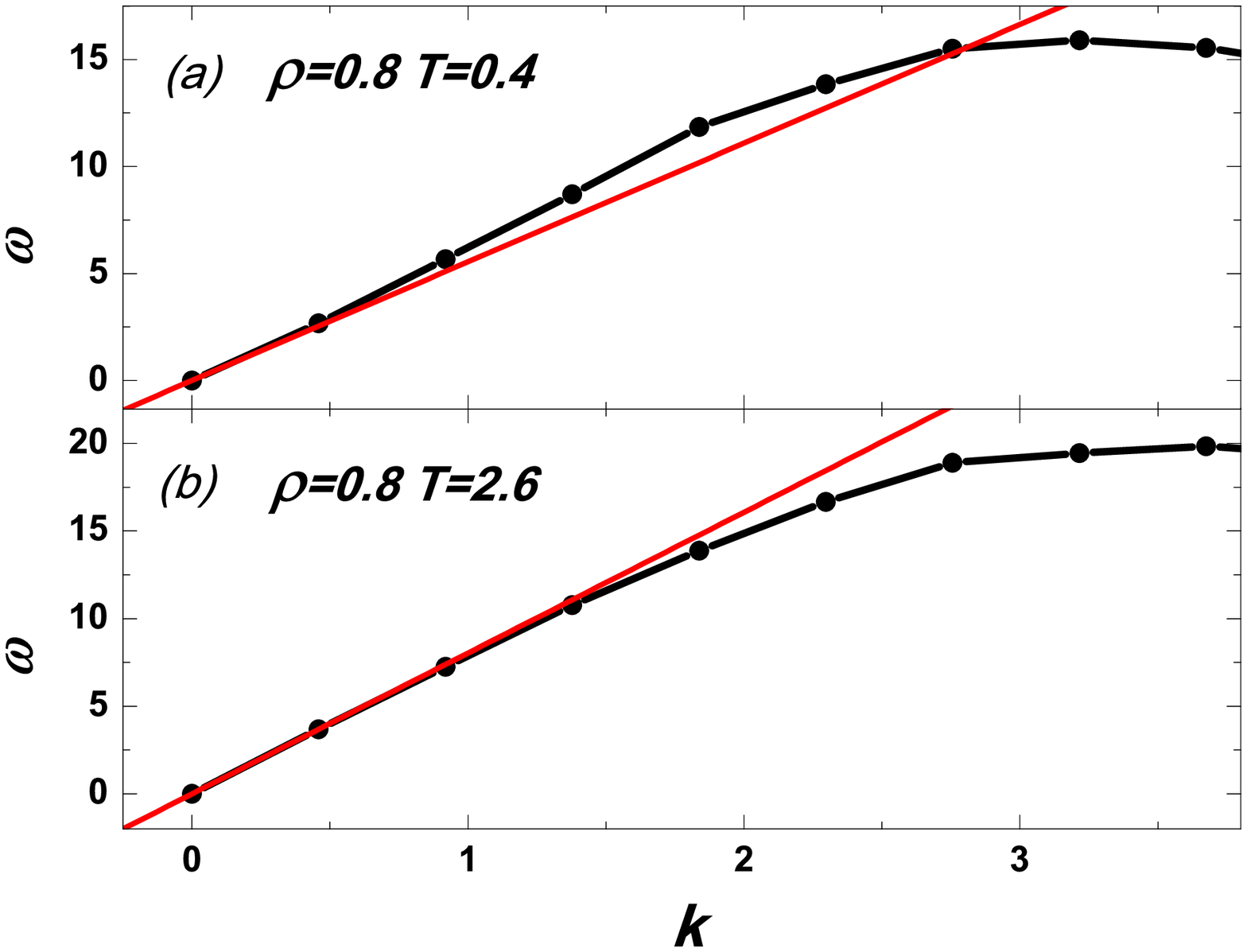}

\caption{\label{fig:fig3} (Color online) Spectra of longitudinal
excitations in system 1 at $\rho=0.8$ and (a) $T=0.4$ and (b)
T=2.6. The straight lines are $c_s \cdot k$, where $c_s$ is adiabatic speed of sound.}
\end{figure}

Fig.~\ref{fig:fig3} (a) and (b) show the dispersion curves for the
system 1 at density $\rho=0.8$ and two temperatures - below and
above the Frenkel line obtained by vacfs and heat capacity
criteria. At low temperature $T=0.4$ a clear positive dispersion
is observed. If the temperature is increased the excitation
frequences approach the line $c_s \cdot k$. Finally the PSD
disappears. However, the point of PSD loss is rather difficult to
be detected because it becomes close to the level of computational
errors.

\begin{figure}
\includegraphics[width=7cm, height=7cm]{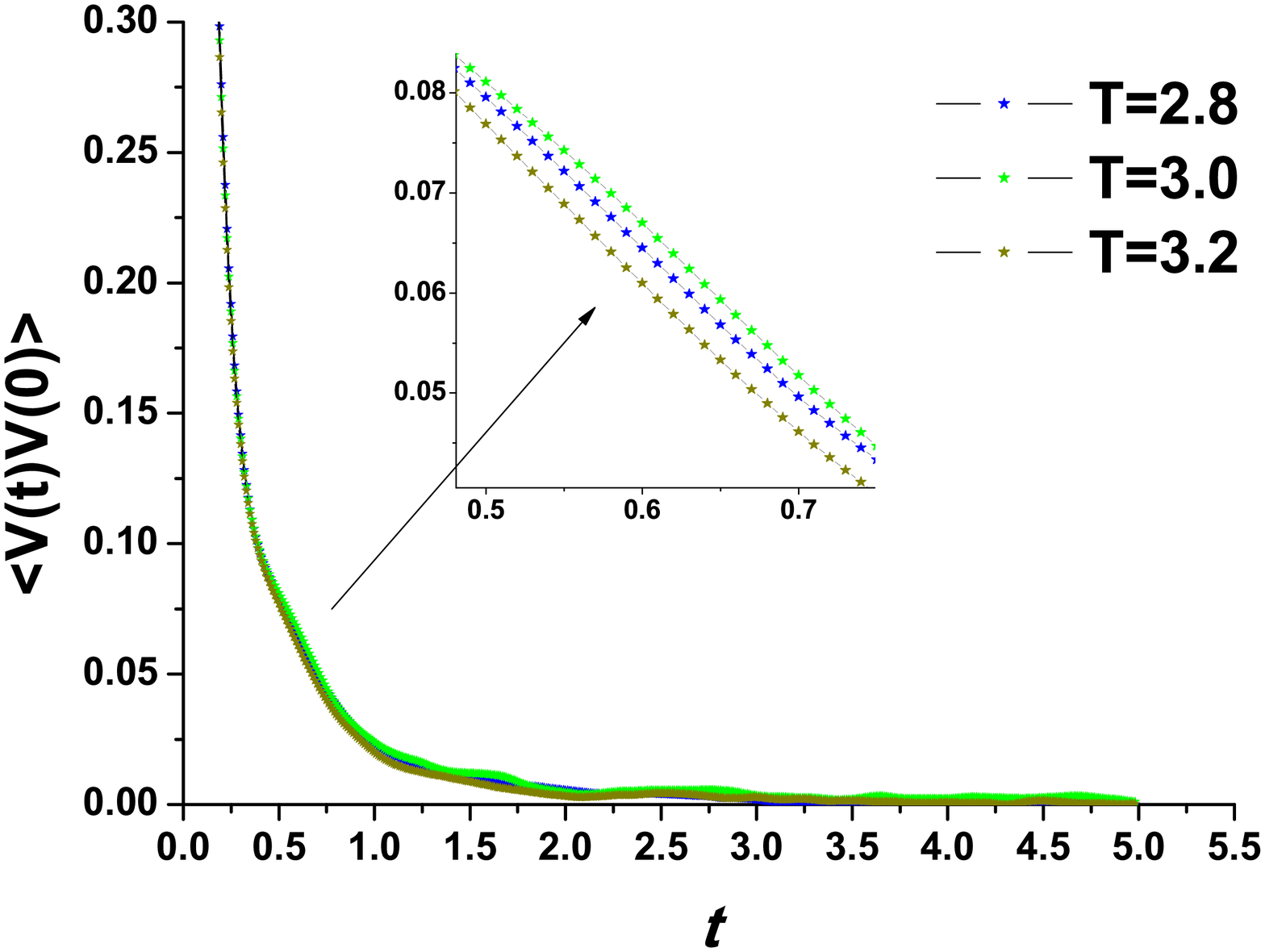}

\caption{\label{fig:fig4a} (Color online) Vacfs of the system in
the vicinity of the high temperature branch of PSD loss at
$\rho=0.55$. The PSD disappears at this density at $T=3.0$.}
\end{figure}

Like in the case of vacfs we observe some kind of reversible
behavior of PSD. There is a density region where if the fluid is
heated isochorically PSD disappears then appears again and
disappears for the second time at higher temperatures. In case of
density $\rho=0.8$ the temperature of the first PSD loss is
$T_1=1.6$ and the second one is $T_2=2.2$. Indeed, the errors in
PSD calculations are quite large and one can relate the second PSD
appearance to the computational uncertainties. However, taking
into account complex behavior of vacfs one can suppose that PSD
also demonstrate sophisticated behavior. Moreover, one can note
that two branches of PSD loss at low densities correspond to two
effects on vacfs. The lower temperature branch is consistent with
the loss of oscillations of vacfs. Apparently at the temperatures
of the second branch of PSD loss vacfs are monotonous both at
lower and higher temperatures, however, non-monotonous behavior of
intensity of vacfs is observed, i.e. the curve of vacf for $T=2.8$
is located below the one for $T=3.0$ but above the curve for
$T=3.2$ (see Fig. ~\ref{fig:fig4a}).


Frenkel lines of the system 1 obtained from all three criteria are
summarized in Fig.~\ref{fig:fig4}. One can see that at the high
density regime ($\rho > 0.75$) all three curves perfectly match.
In our previous publication the same match was found for
Lennard-Jones and soft spheres systems \cite{frprl}. The high
density regime corresponds to the major role of the $d$ length
scale, i.e. $\left(\frac{d}{r} \right)^n$ term of the potentail.
Therefore this result looks to be consistent with our previous
works.

At low densities the situation is more sophisticated. One can see
that the curves from $c_V=2k_B$, the first loss of PSD and
disappearance of oscillations of vacfs are located close to each
other and can be considered as matching. However, the second
temperature of PSD loss is substantionally higher then all these
lines. Moreover, in the region where vacfs demonstrate two points
of loss of oscillations the second branch of the lines from vacfs
is located below the common curve from three criteria.

\begin{figure}
\includegraphics[width=7cm, height=7cm]{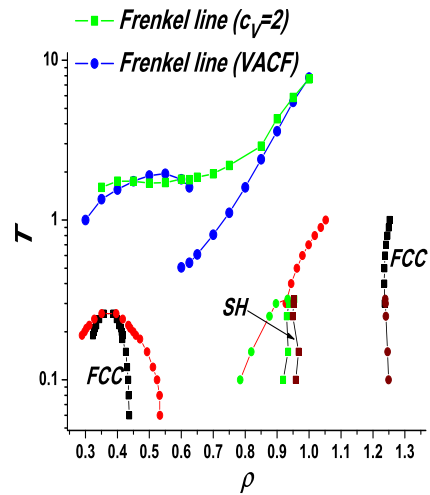}

\caption{\label{fig:fig5} (Color online) Frenkel line of the
system 2 placed in the phase diagram.}
\end{figure}

From the discussion above one can see that in a purely repulsive
core-softened system the Frenkel line behavior becomes very
complicated. It is interesting to analyze the behavior of the
Frenkel line for the systems with both repulsive and attractive
parts of the potential. System 2 is an example of a core-softened
system with both repulsive and attractive forces. The Frenkel
lines of the system 2 from vacf and $c_v=2k_B$ criteria are shown
in Fig.~\ref{fig:fig5}. One can see that like in the case of
system 1 the lines from vacf criterion split into two branches. At
low densities the upper branch of the line from vacfs is in good
coincidence with the line from $c_v$. At higher densities all
lines merge. One can conclude from this picture that the
attractive part of the potential does not change the qualitative
behavior of the Frenkel line.

\bigskip

In conclusion, we report a detailed study of dynamical crossover
line in supercritical regime of core-softened fluids. We use three
criteria - vacfs, $c_V$ and PSD ones and we find good agreement
between them. At low densities the Frenkel line is flat: the
temperature is almost independent on density, while at higher
densities it rapidly grows up. Similar behavior was recently
observed for the Frenkel line of water \cite{fr-water}. Besides we
observe that the crossover lines from vacfs and PSD split into two
pseudo branches at low densities. The phenomena of the appearance
of the additional branches of the Frenkel line is not observed in
simple systems and is caused by quasi-binary nature of the
core-softened fluids.
\bigskip

\begin{acknowledgments}
Eu. G. thanks the Joint Supercomputing Center of Russian Academy
of Science and Yu. F. thanks the Russian Scientific Center at
Kurchatov Institute and for computational facilities. The work was
supported by Russian Science Foundation (Grant No 14-22-00093).
\end{acknowledgments}


\end{document}